\documentclass[aps,pra,reprint,amsmath,amssymb,superscriptaddress,nofootinbib,longbibliography]{revtex4-2}

\usepackage{graphicx}
\usepackage[utf8]{inputenc}
\usepackage[T1]{fontenc}
\usepackage{xcolor}
\usepackage{soul}
\usepackage{amssymb}
\usepackage{bm}

\begin{document}

\title{Tuning magnetostructural phase transition in CoMn$_{0.88}$Cu$_{0.12}$Ge
by application of hydrostatic pressure}

\author{Aleksandra Deptuch}
\affiliation{Institute of Nuclear Physics Polish Academy of
Sciences, Radzikowskiego 152, PL-31-342 Krak\'ow, Poland}
\author{Ryszard Duraj}
\affiliation{Institute of Physics, Cracow University of Technology, Podchor\k{a}\.zych 1,
PL-30-084 Krak\'ow, Poland}
\author{Andrzej Szytu\l{}a}
\affiliation{M.~Smoluchowski Institute of Physics, Jagiellonian University,
prof. Stanis\l{}awa \L{}ojasiewicza 11, PL-30-348 Krak\'ow, Poland}
\author{Bogus\l{}aw Penc}
\affiliation{M.~Smoluchowski Institute of Physics, Jagiellonian University,
prof. Stanis\l{}awa \L{}ojasiewicza 11, PL-30-348 Krak\'ow, Poland}
\author{Ewa Juszy\'nska-Ga\l{}\k{a}zka}
\affiliation{Institute of Nuclear Physics Polish Academy of
Sciences, Radzikowskiego 152, PL-31-342 Krak\'ow, Poland}
\affiliation{Department of Chemistry, Graduate School of Science, Osaka University,
1-1 Machikaneyama, Toyonaka, 560-0043, Osaka, Japan}
\author{Stanis\l{}aw Baran}
\email{stanislaw.baran@uj.edu.pl}
\affiliation{M.~Smoluchowski Institute of Physics, Jagiellonian University,
prof. Stanis\l{}awa \L{}ojasiewicza 11, PL-30-348 Krak\'ow, Poland}

\date{\today}

\begin{abstract}

Structural and magnetic properties of the CoMn$_{0.88}$Cu$_{0.12}$Ge compound have been
investigated using X-ray diffraction (XRD) and differential scanning calorimetry (DSC)
accompanied with investigation of magnetic properties under applied hydrostatic pressure up to 12~kbar.
It has been found that the orthorhombic crystal structure of the TiNiSi-type, which is
dominant at low temperatures, turns into the hexagonal Ni$_2$In-type one, slightly above 200~K.
The structural transition is of the first-order type as confirmed by presence of distinct
thermal hysteresis exceeding 20~K. The ac magnetic measurements indicate a ferromagnetic
order below the Curie temperature $T_C$ of 238~K, followed by additional anomaly
at lower temperatures -- the latter one being related to the structural transition.
Application of hydrostatic pressure leads to temperature separation of the purely
magnetic transition, whose Curie temperature $T_C$ slowly increases with applied pressure,
and the magnetostructural transition characterized by critical temperature
showing much rapid decrease with increasing pressure.
Comparison of the DSC data for the investigated compound and the isostructural
CoMn$_{0.95}$Cu$_{0.05}$Ge one has made it possible to determine both the structural
and magnetic components of entropy change.

\bigskip

\noindent \textbf{keywords}: intermetallic compounds, X-ray diffraction (XRD), ferromagnetic materials,
magnetic properties, pressure-temperature phase diagram

\end{abstract}

\maketitle

\section{Introduction}
\label{intro}

Magnetostructural coupling in the MM'Ge compounds, where M and M' are transition 3d elements,
has been extensively investigated in recent years. The coupling is strictly related to the
structural phase transition between the low-temperature orthorhombic martensite (M) phase and
the high-temperature hexagonal austenite (A) one. %
Stoichiometric \mbox{CoMnGe} has an orthorhombic crystal structure of the TiNiSi-type (space group $Pnma$)
at room temperature. The structure undergoes a martensite structural transformation into the high-temperature
hexagonal Ni$_2$In-type structure (space group $P6_3/mmc$) around $T_s \approx 500$~K. The compound shows also
a second order magnetic transition from para- to ferromagnetic state at the Curie temperature
$T_C \approx 350$~K~\cite{johnson1975diffusionless,PAL201922,PAL2020109036}. Doping of substitutional and/or interstitial atoms as well as
introducing metal vacancies leads to gradual reduction of the difference between $T_s$ and $T_C$, finally resulting
in appearance of a magnetostructural phase transition where both the structural and magnetic transitions
occur at the same temperature. This magnetostructural transition is accompanied with large magnetocaloric
effect~\cite{HAMER20093535,Trung_App_Phys_Lett_96}.

Especially interesting are compounds with the Mn atoms partially substituted by the Cu atoms, as the
neutron diffraction studies indicate that the Mn sublattice is a dominant source of magnetism in this class
of materials~\cite{NIZIOL1982281,SZYTULA1981176}. Until now, the X-ray diffraction data at room temperature
as well as the calorimetric and magnetic properties in function of temperature for the CoMn$_{1-x}$Cu$_x$Ge
system with $x=0-0.15$~\cite{PAL201922,PAL2020109036} and $x=0.08-0.1$~\cite{Tapas_Appl_Phys_Lett_101} have
been reported. On the basis of these data, the structural and magnetic $(T,x)$ phase diagram has been
determined (see Fig.~6b in Ref.~\cite{PAL201922} and Fig.~5 in Ref.~\cite{PAL2020109036}). The temperature
of the structural phase transition between the high-temperature hexagonal phase and the low-temperature
orthorhombic one has been reported to decrease with increasing Cu content $x$ from about 500~K for
$x=0$ down to about 100~K for $x=0.15$, while the Curie temperature decreases from about 350~K for $x=0$
down to about 250~K for $x=0.15$~\cite{PAL201922,PAL2020109036}. The X-ray diffraction data, collected
at room temperature, show dominant contribution arising from the orthorhombic structure for $x=0.08$, while
the hexagonal structure being dominant for $x>0.08$~\cite{Tapas_Appl_Phys_Lett_101}. A giant magnetocaloric
effect has been reported for the Cu concentration range where the first-order magnetostructural transition occurs,
i.e. the temperatures of the structural and magnetic transitions coincide. According to Refs.~\cite{PAL201922,PAL2020109036},
the magnetostructural coupling has been found for $0.09 \leq x \leq 0.11$, while partial coupling has been reported
for $x=0.12$. Ref.~\cite{Tapas_Appl_Phys_Lett_101} reports magnetostructural coupling even for $x=0.08$ and 0.085.
Magnetic entropy change, under magnetic flux change of 0-5~T, has been reported to reach around
50~J$\cdot$kg$^{-1}\cdot$K$^{-1}$ for $x=0.11$~\cite{PAL201922,PAL2020109036} or for $x=0.08$ and
0.085~\cite{Tapas_Appl_Phys_Lett_101}.

In order to better understand the complex magnetostructural properties of the CoMn$_{1-x}$Cu$_x$Ge
system, in this work we report the results of X-ray diffraction (XRD) and differential scanning calorimetry (DSC)
measurements accompanied with investigation of magnetic properties under applied hydrostatic pressure up to 12~kbar,
as performed for CoMn$_{0.88}$Cu$_{0.12}$Ge, i.e. the composition showing partial magnetostructural coupling, according
to the $(T,x)$ phase diagram presented in Refs.~\cite{PAL201922,PAL2020109036}. On the basis of our data, the
$(p,T)$ phase diagram has been determined. In addition, the comparison of the DSC data for the investigated compound
and the isostructural CoMn$_{0.95}$Cu$_{0.05}$Ge one has made it possible to determine both the structural and
magnetic components of entropy change.

\section{Experimental details and results}

\subsection{Sample synthesis}

Polycrystalline sample of CoMn$_{0.88}$Cu$_{0.12}$Ge has been obtained by arc melting of the constituent elements
(purity better than 99.9 wt. \%) under argon atmosphere, followed by annealing
in vacuum ($\sim$$10^{-3}$ mbar) for 5~days at 1123~K (850~$^{\circ}$C) and subsequent furnace cooling down to room
temperature.

\subsection{Crystal structure}

\begin{figure}[!ht]
\begin{center}
\includegraphics[bb=14 14 1007 1291, width=\columnwidth]
	{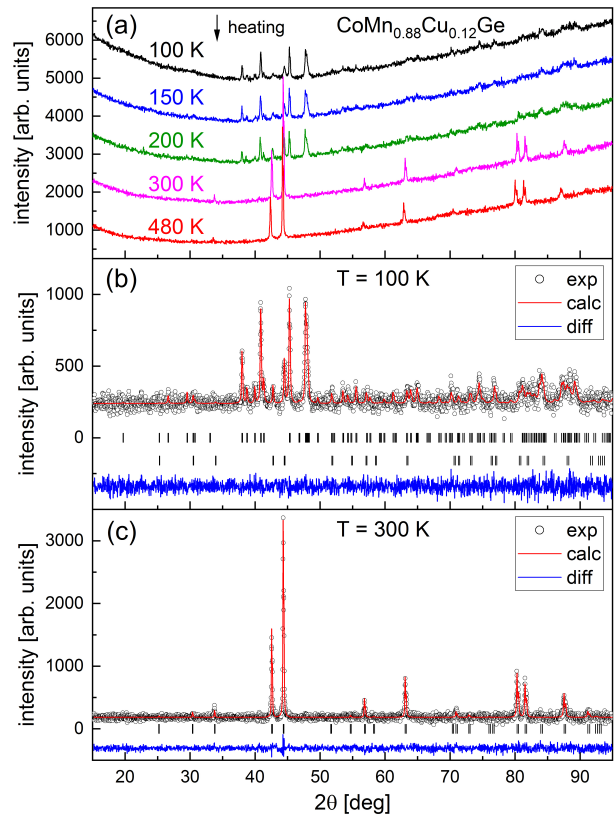}
\end{center}
\caption{\label{fig:XRD_patterns}
XRD patterns of CoMn$_{0.88}$Cu$_{0.12}$Ge: (a) Reprezentative patterns collected at different temperatures
on heating. (b) Pattern collected at 100 K with the results of Rietveld refinement (background subtracted).
The first and second row of vertical bars indicates the peak positions originating from the orthorhombic
phase (88.8(9)~wt.~\%) and hexagonal one (11.2(8)~wt.~\%), respectively. (c) Pattern collected at 300~K
with the results of Rietveld refinement (background subtracted). The vertical bars indicate the peak
positions from the hexagonal phase.} 
\end{figure}

Thermal evolution of the powder X-ray diffraction pattern for CoMn$_{0.88}$Cu$_{0.12}$Ge clearly
indicates a temperature-induced structural transition -- see Fig.~\ref{fig:XRD_patterns}a. In purpose
to determine the exact crystal structures, a Rietveld refinement of the X-ray diffraction data has been
performed using the FullProf software~\cite{RODRIGUEZCARVAJAL199355}. Figs.~\ref{fig:XRD_patterns}b
and \ref{fig:XRD_patterns}c show the XRD patterns collected at 100 and 300~K, respectively, together
with the results of the Rietveld refinement. The experimental data indicate that CoMn$_{0.88}$Cu$_{0.12}$Ge
at 100~K (Fig.~\ref{fig:XRD_patterns}b)
is a mixture of two crystal phases, namely, the orthorhombic one (88.8 wt. \%) of the TiNiSi-type
(space group $Pnma$, No. 62) in which each element occupy the $4c$ site $(x,\frac{1}{4},z)$
with different values of the $x$ and $z$ parameters, and the hexagonal phase (11.2 wt. \%) of the
Ni$_2$In-type (space group $P6_3/mmc$, No. 194) with the Co atoms occupying the $2d$ site
$(\frac{1}{3},\frac{2}{3},\frac{3}{4})$, while the Mn and Cu atoms occupying the same $2a$ site
$(0,0,0)$ and the Ge atoms located in the $2c$ site $(\frac{1}{3},\frac{2}{3},\frac{1}{4})$.
CoMn$_{0.88}$Cu$_{0.12}$Ge at 300~K (see Fig.~\ref{fig:XRD_patterns}c) is single-phase,
consisting of the hexagonal phase solely.

\begin{figure}[!ht]
\begin{center}
\includegraphics[bb=14 14 411 865, width=\columnwidth]
	{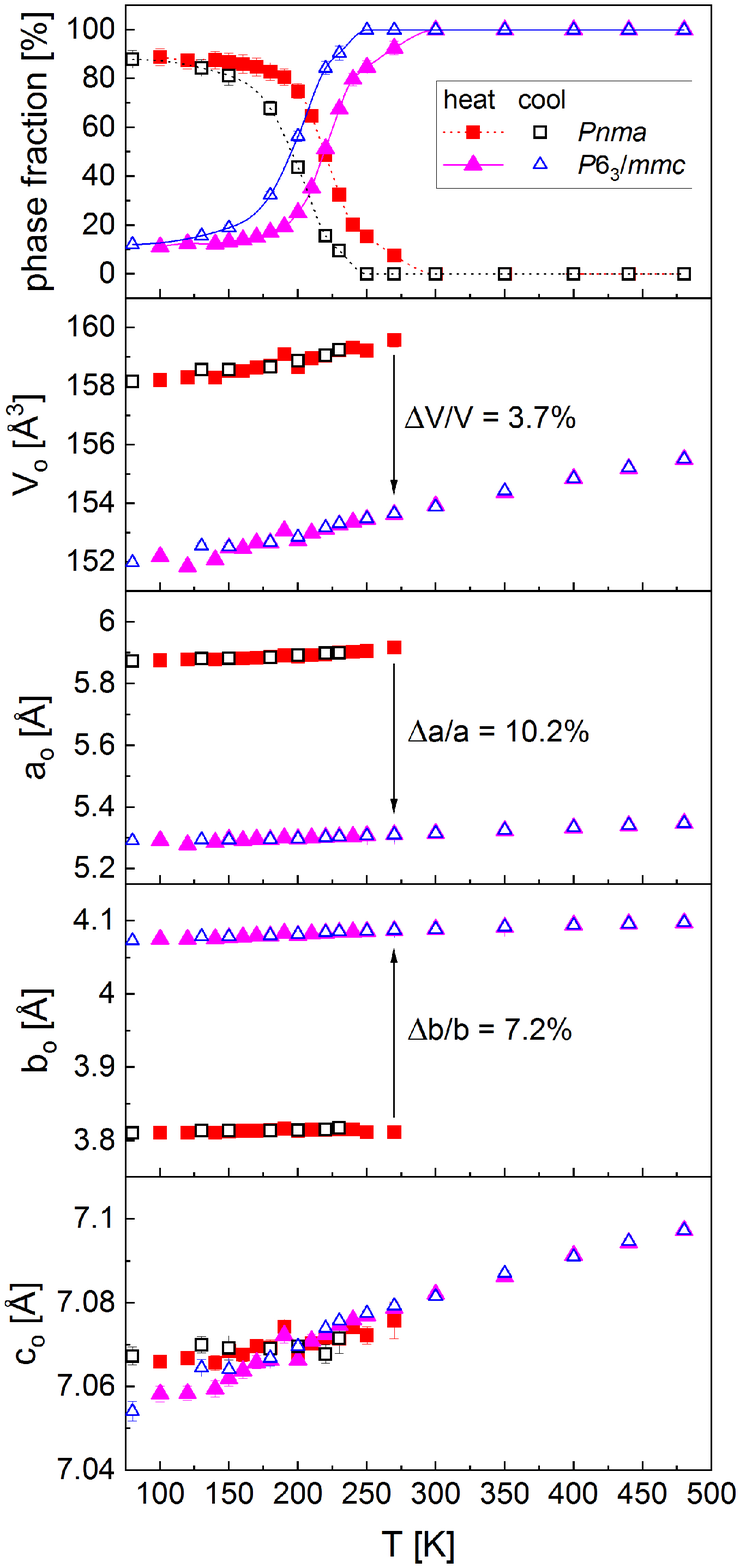}
\end{center}
\caption{\label{fig:unit_cell_params}
Phase fraction (weight percentage), unit cell volume and lattice constants vs. temperature for
CoMn$_{0.88}$Cu$_{0.12}$Ge. Unit cell parameters of the hexagonal phase have been recalculated
to those of the orthorhombic phase using the following relations: $a_o=c_h$, $b_o=a_h$,
$c_o=\sqrt{3}a_h$ and $V_o=2V_h$.}
\end{figure}

Figure~\ref{fig:unit_cell_params} shows thermal evolution of: fractions of the orthorhombic and hexagonal
phases, lattice parameters $a$, $b$ and $c$ as well as unit cell volume $V$, as determined from
X-ray powder diffraction. The orthorhombic phase, which is dominant at low temperatures, undergoes
an inverse martensitic transition to the hexagonal phase at 219~K, while the high-temperature
hexagonal phase transforms into the orthorhombic one on cooling (martensitic transition) at 193~K.
The transition temperatures have been defined as inflection points in the phase fraction vs. temperature
curves. A comparison of the lattice parameters characterizing the orthorhombic and hexagonal phases
can be done considering the relationships: $a_o=c_h$, $b_o=a_h$, $c_o=\sqrt{3}a_h$ and $V_o=2V_h$.
A first order character of the structural transition is confirmed by a large volume contraction
$\frac{\Delta V}{V} = 3.7$~\%, step change of selected lattice parameters, namely,
$\frac{\Delta a}{a_o} = 10.2$~\% and $\frac{\Delta b}{b_o} = 7.2$~\% as well as
thermal hysteresis visible in the phase fraction vs. temperature curves.

\begin{table*}
\caption{\label{tab:lattice}
Lattice parameters $a$, $b$ and $c$ and the unit cell volume $V$ of the orthorhombic and hexagonal phases
together with the phase fraction of the hexagonal phase, as determined from the X-ray diffraction data
collected at 100, 300 and 480~K.}
\begin{footnotesize}
\begin{tabular*}{\textwidth}{@{\extracolsep{\fill}}ccccccccc}
T [K] & $a_o$ [\AA{}] & $b_o$ [\AA{}] & $c_o$ [\AA{}] & $V_o$ [\AA{}$^3$] & $a_h$ [\AA{}] & $c_h$ [\AA{}] & $V_h$ [\AA{}$^3$] & hex [\%]\\ \hline
100 & 5.876(2) & 3.8104(8) & 7.066(2) & 158.22(7) & 4.075(1) & 5.291(4) & 76.10(6) & 11.2(8)\\
300 & - & - & - & - & 4.0889(4) & 5.3158(5) & 76.97(2) & 100\\
480 & - & - & - & - & 4.0976(4) & 5.3472(6) & 77.75(2) & 100\\
\end{tabular*}
\end{footnotesize}
\end{table*}

The values of refined lattice parameters and the unit cell volumes at 100, 300 and 480~K are listed in
Table~\ref{tab:lattice}. The data presented in Fig.~\ref{fig:unit_cell_params} and Table~\ref{tab:lattice}
indicate almost linear thermal expansion of lattice parameters of the hexagonal phase.

\subsection{Thermal properties}

\begin{figure}[!ht]
\begin{center}
\includegraphics[bb=14 14 808 865, width=\columnwidth]
        {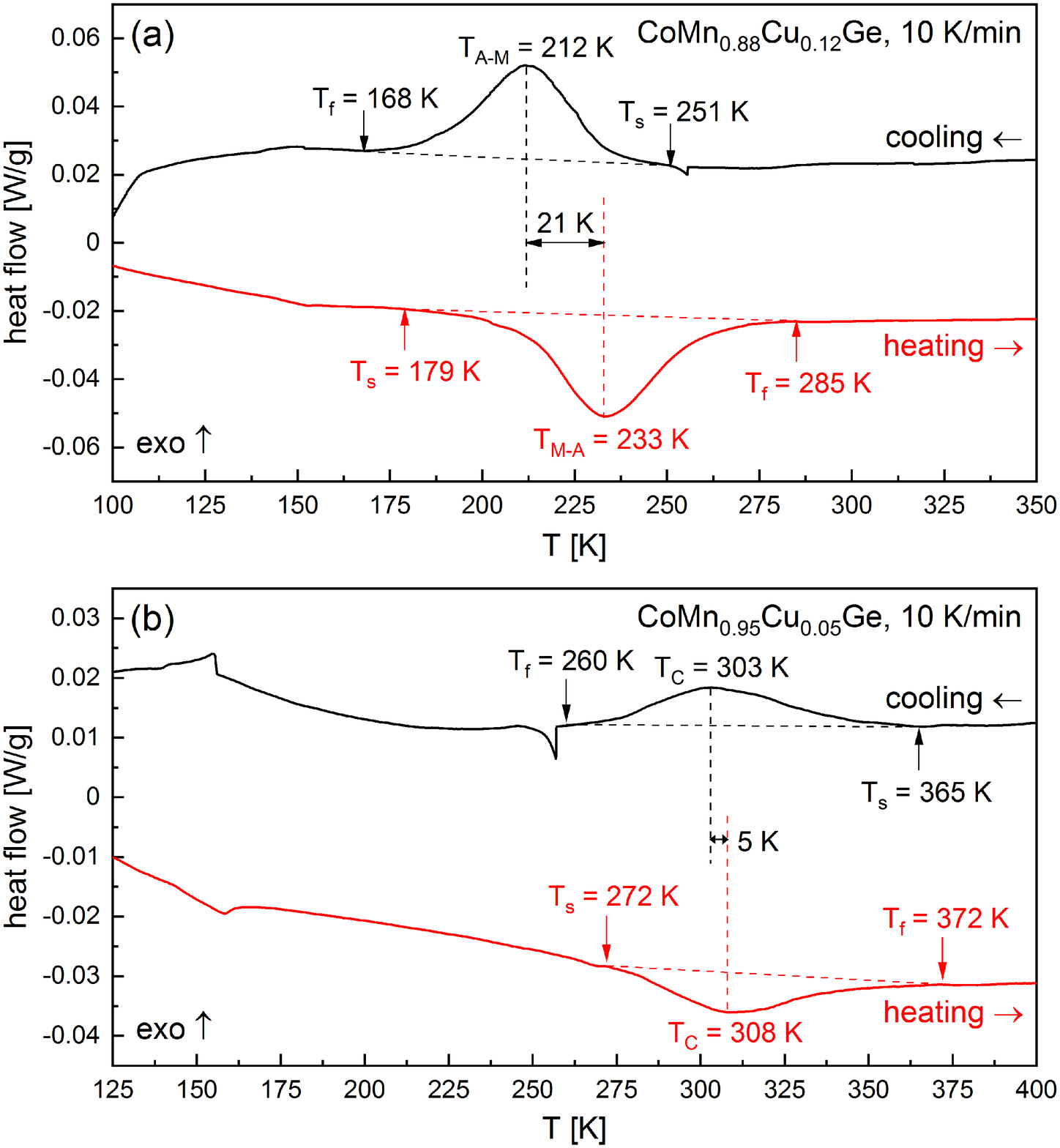}
\end{center}
\caption{\label{fig:DSC}
DSC thermograms of (a) CoMn$_{0.88}$Cu$_{0.12}$Ge and (b) CoMn$_{0.95}$Cu$_{0.05}$Ge at cooling and subsequent
heating with the rate of 10~K/min. The $T_s$ and $T_f$ symbols refer respectively to the start and finish of
the structural transition during cooling and heating. In (a), the characteristic temperatures of transitions
from austenite to martensite ($T_{A-M}$) and from martensite to austenite ($T_{M-A}$) are defined as the peak
positions. In (b), $T_C$ denotes the Curie temperature, while the small anomalies at 160 and 260~K correspond to
the magnetic transitions in the impurity hexagonal phase which is observed for the samples with large Cu
content (the XRD analysis of the sample indicates about 9.6~wt.~\% of the impurity phase).}
\end{figure}

Differential scanning calorimetry (DSC) measurements were carried in the temperature range 100-350~K with
the temperature rate of 10~K/min in DSC 2500 (TA Instruments) calorimeter. The results of the DSC measurements
are shown in Fig.~\ref{fig:DSC}a.

\begin{table}
\caption{\label{tab:DSC_CoMn0.88Cu0.12Ge}
Enthalpy ($\Delta H$) and total entropy ($\Delta S$) changes associated with the structural phase transition
in CoMn$_{1-x}$Cu$_x$Ge, $x$ = 0.05 and 0.12, as determined from the DSC data collected during heating and cooling.}
\begin{footnotesize}
\begin{tabular*}{\columnwidth}{@{\extracolsep{\fill}}ccccc}
 & $\Delta H$ & $\Delta H$ & $\Delta S$ & $\Delta S$ \\
compound & [J/g] & [kJ/mol] & [J/(kg$\cdot$K)] & [J/(mol$\cdot$K)] \\ \hline
$x$ = 0.05, cooling & 1.79 & 0.34 & 5.83 & 1.09 \\
$x$ = 0.05, heating & 1.79 & 0.34 & 5.72 & 1.07 \\
$x$ = 0.12, cooling & 4.97 & 0.93 & 23.7 & 4.44 \\
$x$ = 0.12, heating & 5.95 & 1.12 & 25.5 & 4.79 \\
\end{tabular*}
\end{footnotesize}
\end{table}

A difference between the peak temperatures obtanined during heating and cooling is marked by vertical
dashed lines in Fig.~\ref{fig:DSC}a. A noticeable thermal hysteresis of 21~K indicates
first order character of the structural phase transition. Determined values of the enthalpy ($\Delta H$) and
total entropy ($\Delta S$) changes are given in Table~\ref{tab:DSC_CoMn0.88Cu0.12Ge}.

\subsection{Magnetic properties}

High pressure magnetic measurements were carried out in the temperature range 80-400~K under fully hydrostatic
conditions. Helium was used as a pressure-transmitting medium. The pressure cell (made of beryllium bronze)
was connected to the UNIPRESS GCA-10 three stage gas compressor by a manganin gauge placed in the highest
pressure stage of the compressor. Temperature was controlled by a temperature controller equipped with a
thermocouple placed directly at the sample position. The ac magnetic susceptibility measurements were carried
out in a weak magnetic field of 1~mT (10~Oe) amplitude and frequency of 300~Hz. The voltage induced in the pick-up
coils was measured by a lock-in amplifier.

\begin{figure}[!ht]
\begin{center}
\includegraphics[bb=14 14 865 1007, width=\columnwidth]
        {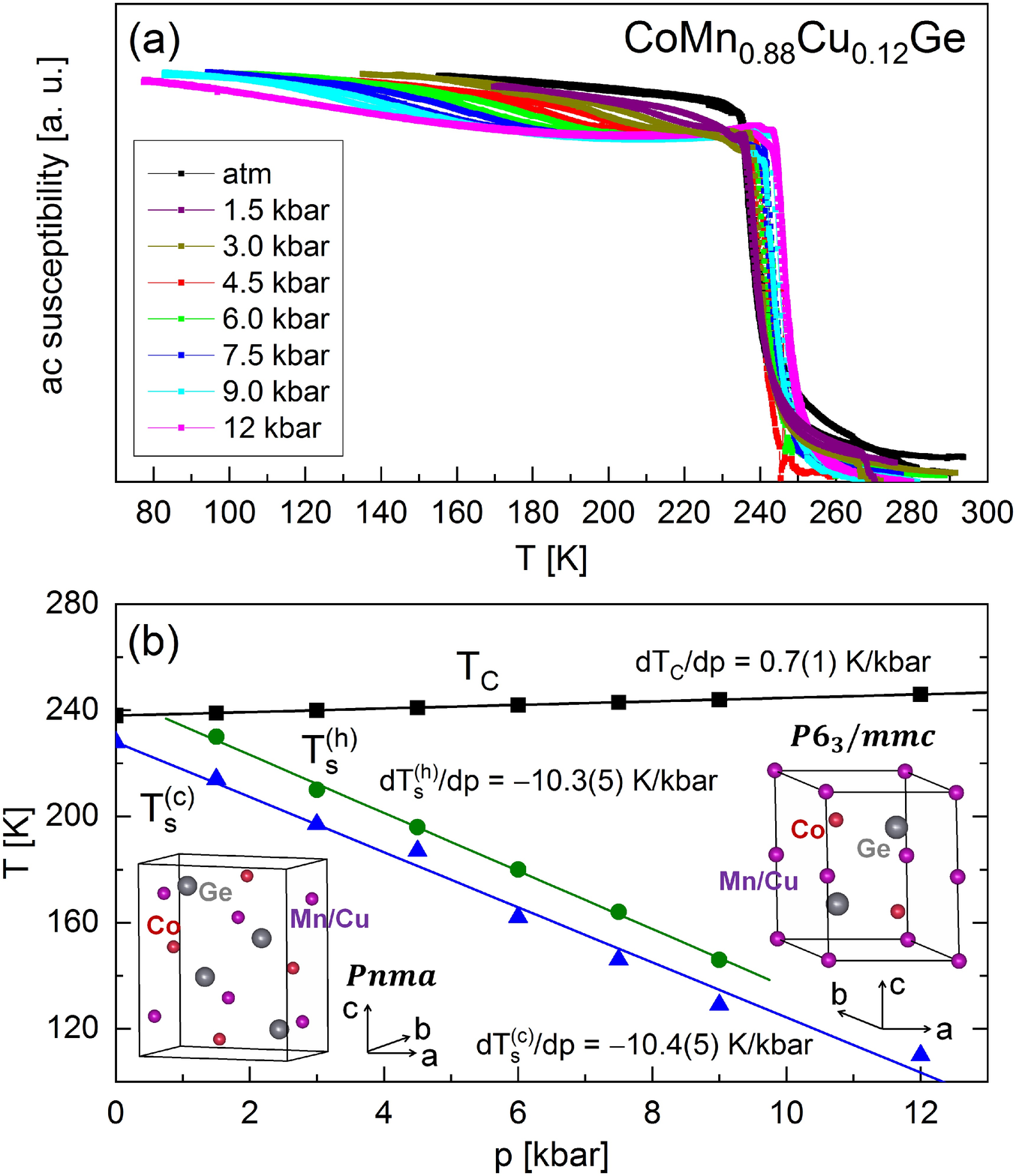}
\end{center}
\caption{\label{fig:ac_magn_susc_and_p-T_diagram}
(a) Temperature dependence of the ac magnetic susceptibility of CoMn$_{0.88}$Cu$_{0.12}$Ge at selected hydrostatic pressures up
to 12~kbar. (b)~Magnetostructural $(p,T)$ phase diagram of CoMn$_{0.88}$Cu$_{0.12}$Ge. $T_C$ and $T_s$ denote the Curie and
structural phase transitions, respectively, while (h) and (c) indices refer to heating and cooling. The error bars are smaller
than the plot symbols.}
\end{figure}

Fig.~\ref{fig:ac_magn_susc_and_p-T_diagram}a shows temperature dependence of the ac magnetic susceptibility $\chi$ for CoMn$_{0.88}$Cu$_{0.12}$Ge
during heating and cooling within the temperature range 80-360~K and selected hydrostatic pressures up to 12~kbar.
At ambient pressure, a sharp increase of ac magnetic susceptibility with deacreasing temperature, cha\-rac\-te\-ris\-tic of
transition from para- to ferromagnetic state, is vi\-si\-ble at $T_C=238$~K. The Curie temperature has been defined as an inflection
point in the $\chi(T)$ curve. A slow increase of $T_C$ with increasing pressure is observed.

A distinct thermal hysteresis in the $\chi(T)$ curves is noticeable below the respective Curie points. An anomaly, defined as a
second inflection point in the $\chi(T)$ curve, can be attributed to the magnetostructural transition associated with the
(inverse) martensitic transition in CoMn$_{0.88}$Cu$_{0.12}$Ge. The temperature of anomaly ($T_s$) decreases with
increasing pressure and shows thermal hysteresis of almost 20~K.

On the basis of the collected ac magnetic measurements, the magnetostructural pressure-temperature $(p,T)$ phase diagram
has been determined (see Fig.~\ref{fig:ac_magn_susc_and_p-T_diagram}b). The Curie temperature is found to increase slowly with increasing
pressure ($\textrm{d}T_C/\textrm{d}p=0.7(1)$~K/kbar). The characteristic temperature of the magnetostructural
transition is lower than the Curie one, and decreases with increasing pressure with much higher rate
($\textrm{d}T_s/\textrm{d}p=-10.4(5)$~K/kbar). The first order character of the latter transition is confirmed by
presence of distinct thermal hysteresis of $T_s$.

\section{Discussion}

The X-ray powder diffraction data show that the low-temperature orthorhombic crystal structure (TiNiSi-type) of
CoMn$_{0.88}$Cu$_{0.12}$Ge undergoes an (inverse) martensitic transition into the hexagonal one (Ni$_2$In-type) with
increasing temperature. Distinct jumps in the unit cell volume $V$ as well as the lattice parameters $a$ and $b$
(see Fig.~\ref{fig:unit_cell_params}) are characteristic of the first order-type transition. The first order of
the structural transition is confimed by thermal hysteresis found in both the DSC (Fig.~\ref{fig:DSC}a)
and ac magnetic data (Fig.~\ref{fig:ac_magn_susc_and_p-T_diagram}a). The magnetic
data, taken at ambient pressure, reveal that further increase od temperature leads to another phase transition
from the low-temperature ferromagnetic state to the high-temperature paramagnetic one. Proximity of two transition
temperatures leads to partial coupling of the structural and magnetic transitions at ambient pressure.

The pressure-temperature $(p,T)$ phase diagram, determined on the basis of the ac magnetic measurements under
hydrostatic pressures up to 12~kbar (see Fig.~\ref{fig:ac_magn_susc_and_p-T_diagram}b), shows that increasing pressure breaks the
partial coupling of the structural and magnetic transitions, as the Curie temperature increases slowly with increasing
pressure ($\textrm{d}T_C/\textrm{d}p=0.7(1)$~K/kbar) while the temperature of the second transition (having
magnetostructural character) decreases with increasing pressure with much higher rate
($\textrm{d}T_s/\textrm{d}p=-10.4(5)$~K/kbar). A similar effect can be obtained by applying a chemical pressure --
compare with the $(T,x)$ phase diagram (Fig.~6b in Ref.~\cite{PAL201922} and Fig.~5 in Ref.~\cite{PAL2020109036}),
where $x$ denotes the copper content in CoMn$_{1-x}$Cu$_x$Ge. The chemical pressure increases with increasing
copper content as the copper radius ($r_{Cu} \approx 1.28$~\AA) is smaller than that of manganese ($r_{Mn} \approx 1.30$~\AA)~\cite{Pearson1972}.

Total entropy change $\Delta S_T$, associated with a magnetostructural transition, is a sum of the magnetic entropy
change $\Delta S_m$ and entropy difference $\Delta S_s$
between two different crystallographic polymorphs, i.e. $\Delta S_T = \Delta S_m + \Delta S_s$~\cite{GSCHNEIDNER2012572}.
In order to determine individual components of the sum for CoMn$_{0.88}$Cu$_{0.12}$Ge, additional DSC measurements
have been performed for the isostructural CoMn$_{0.95}$Cu$_{0.05}$Ge, where purely magnetic transition occurs
for the low-temperature orthorhombic phase, as the temperature of structural transition is found to be around 100~K higher
(Fig.~6b in Ref.~\cite{PAL201922} and Fig.~5 in Ref.~\cite{PAL2020109036}). The DSC curves, collected for
CoMn$_{0.95}$Cu$_{0.05}$Ge, show broad peaks at 308~K for heating and 303~K for cooling
(see Fig.~\ref{fig:DSC}b). The corresponding magnetic entropy changes equal 5.7~J$\cdot$kg$^{-1}\cdot$K$^{-1}$
for heating and 5.8~J$\cdot$kg$^{-1}\cdot$K$^{-1}$ for cooling, respectively. The magnetic entropy change is not sensitive to small
changes in chemical composition as the value for stoichiometric CoMnGe has been reported to equal
5.8~J$\cdot$kg$^{-1}\cdot$K$^{-1}$ under magnetic flux change of 0-5~T~\cite{LAI201486}. Assuming the same value of magnetic
entropy change (i.e. 5.8~J$\cdot$kg$^{-1}\cdot$K$^{-1}$) for CoMn$_{0.88}$Cu$_{0.12}$Ge and the average total entropy change
of 24.6~J$\cdot$kg$^{-1}\cdot$K$^{-1}$ (see Tab.~\ref{tab:DSC_CoMn0.88Cu0.12Ge}) leads to the estimated value of the
structural component of around 19~J$\cdot$kg$^{-1}\cdot$K$^{-1}$. Therefore, the structural component plays
the dominant role in the total entropy change associated with the magnetostructural transition in CoMn$_{0.88}$Cu$_{0.12}$Ge.

\section{Conclusions}

The current work reports the results of a systematic investigation of structural and magnetic phase transitions in
CoMn$_{0.88}$Cu$_{0.12}$Ge by complementary experimental techniques including X-ray diffraction (XRD) and differential
scanning calorimetry (DSC) accompanied with investigation of magnetic properties under applied hydrostatic pressure up
to 12~kbar. The compound shows at ambient pressure a partial coupling of the structural martensitic transition and
the magnetic transition between the ferro- and paramagnetic states. Increasing pressure breaks the
partial coupling of the structural and magnetic transitions, as the Curie temperature increases slowly with increasing
pressure while the temperature of the second transition (having magnetostructural character) decreases with increasing
pressure with much higher rate. Comparison of the DSC data for CoMn$_{0.88}$Cu$_{0.12}$Ge and CoMn$_{0.95}$Cu$_{0.05}$Ge
reveals that the structural component dominates over the magnetic one in the total entropy change associated with the
magnetostructural transition in CoMn$_{0.88}$Cu$_{0.12}$Ge.

\section*{Acknowledgements}

The research was partially carried out with the equipment purchased thanks to the financial support of the European
Regional Development Fund in the framework of the Polish Innovation Economy Operational Program [contract no.
POIG.02.01.00-12-023/08] and the European Regional Development Fund Operational Program Infrastructure and
Environment [contract no. POIS 13.01.00-00-062/08].

\bibliography{CoMn0.88Cu0.12Ge_magn_under_pressure}

\end{document}